# A Heterogeneous Chiplet Architecture for Accelerating End-to-End Transformer Models

Harsh Sharma*

Washington State University, Pullman, WA, USA, harsh.sharma@wsu.edu

Pratyush Dhingra

Washington State University, Pullman, WA, USA, pratyush.dhingra@wsu.edu

Janardhan Rao Doppa

Washington State University, Pullman, WA, USA, jana.doppa@wsu.edu

Umit Y. Ogras

University of Wisconsin Madison, Madison, WI, USA, uogras@wisc.edu

Partha Pratim Pande

Washington State University, Pullman, WA, USA, pande@wsu.edu

Transformers have revolutionized deep learning and generative modeling, enabling advancements in natural language processing tasks. However, the size of transformer models is increasing continuously, driven by enhanced capabilities across various deep learning tasks. This trend of ever-increasing model size has given rise to new challenges in terms of memory and compute requirements. Conventional computing platforms, including GPUs, suffer from suboptimal performance due to the memory demands imposed by models with millions/billions of parameters. The emerging chiplet-based platforms provide a new avenue for compute- and data-intensive machine learning (ML) applications enabled by a Network-on-Interposer (NoI). However, designing suitable hardware accelerators for executing Transformer inference workloads is challenging due to a wide variety of complex computing kernels in the Transformer architecture. In this paper, we leverage chiplet-based heterogeneous integration (HI) to design a high-performance and energy-efficient multi-chiplet platform to accelerate transformer workloads. We demonstrate that the proposed NoI architecture caters to the data access patterns inherent in a transformer model. The optimized placement of the chiplets and the associated NoI links and routers enable superior performance compared to the state-of-the-art hardware accelerators. The proposed NoI-based architecture demonstrates scalability across varying transformer models and improves latency and energy efficiency by up to 11.8× and 2.36×, respectively when compared with the existing state-of-the-art architecture HAIMA.

CCS CONCEPTS • 2.5D • NLP • Processing-in-memory • Network-on-interposer • Transformer • Chiplet-based architecture

* This work was supported in part by the U.S. National Science Foundation under Grant CSR-2308530, and in part by the Army Research Office under Grant ARO-W911NF-24-1-0240. Authors' addresses: H. Sharma, P. Dhingra, J. R. Doppa, and P. P. Pande, Washington State University, Pullman, WA; emails: harsh.sharma@wsu.edu, pratyush.dhingra@wsu.edu, jana.doppa@wsu.edu, pande@wsu.edu; Umit Y. Ogras, University of Wisconsin-Madison, Department of Electrical and Computer Engineering, Madison, WI, 53706, USA; email: uogras@wisc.edu.

# 1 INTRODUCTION

We are at the cusp of a revolution from deep generative AI models, also referred to as *foundation models (FMs)*, for text (e.g., ChatGPT) and images (e.g., stable diffusion) with hundreds of billions of parameters trained on a massive amount of unlabeled data. In the future, fine-tuning FMs on domain-specific private data (e.g., BloombergGPT for finance, AutoGPT for general-purpose automation) to solve a specific task with high accuracy will emerge as a fundamental challenge. The current FMs employ Transformer architectures due to their widespread success in applications, including natural language processing, computer vision, and multi-modal integration [1] [2]. The number of parameters and complexity of transformer models are growing at a rapid pace to meet the application demands, including scale of data and high accuracy. Hence, designing suitable hardware accelerators for Transformer models involves significant computation and storage challenges, particularly when addressing longer contextual information. The main challenge arises from the quadratic relationship between the compute requirements and the input sequence length ($N$). Unlike recurrent neural networks (RNNs), Transformers employ self-attention to process various modalities of data, including text, images, videos, and speech, enabling tokens within sequences to relate to one another effectively. This approach overcomes the vanishing gradient problem in RNNs and allows to effectively leverage long-range sequential dependencies.

Widespread adoption of Transformer models in real-world applications implies a vast demand for designing hardware platforms with adequate computational and storage resources [1] [2] [3]. We must tackle complex data movement, memory hierarchy, and latency challenges while designing suitable hardware architectures. Hence, there is a need for designing large-scale chips with high memory and compute capabilities. Moreover, due to various types of computational kernels involved in Transformer models, we require different types of processing elements, such as Tensor cores, GPUs, DRAMs, and processing-in-memory (PIM)-based accelerators on the same system. Such large-scale and heterogeneous integration increases the area of monolithic chips significantly [4]. One of the major challenges in the silicon industry is the exploding fabrication cost as the size of monolithic chips approaches the reticle limit [4].

Chiplet-based 2.5D systems that integrate multiple smaller chips (chiplets) on a single interposer offer a promising solution for reducing the manufacturing cost of large monolithic chips and enabling large-scale heterogeneous integration in a single system. Chiplets are connected through the network-on-interposer (NoI). In a heterogeneous chiplet-based system, designing a scalable NoI architecture is daunting due to the relatively large physical distances between chiplets, poor technology scaling of electrical wires, and varying data movement patterns depending on the heterogeneity in constituent chiplets. This data movement introduces performance overhead. In addition, managing data movement across different levels of the memory hierarchy is non-trivial, requiring careful placement of chiplets, NoI links, and routers to ensure efficient data access paths. However, current state-of-the-art Transformer accelerators do not consider the challenges associated with the heterogeneous integration (HI) within a single system. We adopt a holistic view by considering the interactions among various computing kernels within transformers to design the HI architecture. We also optimize the system-level performance by choosing a suitable chiplet configuration for each computational kernel. In this paper, we present an NoI backbone design methodology, enabling a heterogeneous chiplet architecture for end-to-end acceleration of Transformer models, and demonstrate significant improvements over existing hardware accelerators.

The major contributions of this paper are as follows:

1. We propose a novel methodology for designing a high-performance and energy-efficient NoI architecture supporting heterogeneous chiplets for accelerating Transformer models.
2. We demonstrate that the proposed NoI architecture caters to the data access patterns inherent in a heterogeneous system. The optimized placement of the chiplets and the associated NoI links and routers enable superior performance compared to the state-of-the-art counterparts.
3. Extensive performance evaluation shows up to 11.8× and 2.36× reduction in latency and energy compared to state-of-the-art architectures.

The rest of the paper is organized as follows. Section II presents the relevant prior work. Section III introduces the overall architecture and explains the associated data flow. Section IV presents the experimental setup and detailed performance evaluation. Finally, Section V concludes the paper by highlighting the salient contributions of this work.

## 2 RELATED WORK

As this work is focused on designing 2.5D accelerators for Transformers, we review the related work in two parts: 2.5D-based manycore architectures and Transformer accelerators.

**2.5D-based manycore architectures:** Both application-specific and general-purpose chiplet architectures have been explored. Design space exploration of 2.5D-based systems considering technology nodes, chiplet-sizes, reducing DRAM costs, manufacturing defects, big-little chiplet paradigm, and multi-link network frequency architectures have been proposed [5] [6] [7] [8] [9]. The NoI paradigm becomes crucial as the communication demand increases with many chiplets integrated on the same substrate [5]. So far, multiple NoI architectures have been proposed in the literature. Most of these architectures are based on conventional multi-hop interconnection architectures, such as mesh or torus [6]. A server-scale application-specific 2.5D architecture called SWAP is proposed for deep learning workloads [10]. SIMBA introduces tiling optimizations on fixed NoI topologies for executing deep models such as ResNet50 [11]. NN-Baton proposes a framework to explore the chiplet design space for convolutional operations [12]. Recent work discusses the advantages of integrating heterogeneous chiplets on the interposer to reduce recurring design costs [13]. A scalable high-performance chiplet-based architecture called HexaMesh has been proposed recently [14]. However, *none of the above 2.5D architecture can be employed off-the-shelf to accelerate Transformer models* as they do not efficiently handle the data-access patterns involving heterogeneous chiplets and computational kernels.

**Transformer accelerators:** PIM is an enabling technology to accelerate deep learning workloads [10]. Several PIM architectures have been proposed for Transformer models. ReTransformer is a ReRAM-based PIM accelerator designed only for the attention kernel and is oblivious to the data access patterns involved in a Transformer model [1]. Additionally, the expected rewrites due to intermediate results during the attention computation within the Transformer model would exceed the write endurance of the ReRAM blocks. We have quantified this limitation in the section 4.4. Xformer is another accelerator that uses ReRAM-SRAM arrays to divide the attention kernel into dynamic and static components, respectively [15]. Xformer maps more frequently updated computation kernels to SRAM arrays. HAIMA is a recently proposed hybrid DRAM-SRAM compute-in-memory architecture [3]. HAIMA accelerates parallel kernels using SRAM and DRAM components. Multiple ASIC designs have been proposed to accelerate the attention kernel only [16] [17] [18]. These ASICS do not accelerate the entire Transformer model. EdgeBERT leverages dynamic voltage-frequency scaling based on the early exit prediction of ALBERT, which is a lightweight Transformer model with reduced memory footprint [19]. A recently proposed monolithic-3D accelerator, called AccelTran prunes activations at runtime [20]. TransCODE is another codesign framework that finds a Transformer-accelerator pair that maximizes the performance objectives within the given user-defined constraints [21]. Newton, FIMDRAM, and McDRAM cascade bit-arithmetic units to do computation near DRAM banks [18] [22]. However, the complicated bit-parallel and bulky buffers incur significant overhead and decrease memory density. Other methodologies, such as TurboTransformer, introduce large area overhead and have limited acceleration due to simpler tiled computation [23]. TransPIM implements the computing kernels in HBM memory stacks with optimized data paths using token sharing in a ring broadcast among memory banks during attention computation. TransPIM uses auxiliary compute units to avoid extra data movement and suffers from latency overhead at each kernel, compromising the overall execution time [2]. Multiple industrial-grade system design explored design techniques to increase compute and on-chip memory within a reticle limit due to the leakage power constraint. A rack-based scaling is employed for large-scale transformer acceleration by Nvidia, AMD, and Intel with DGX A100, EPYC, and Gaudi line of AI processors respectively [24] [25]. Samsung recently proposed Aquabolt-XL, a processing-in-memory (PIM) based 3D-architecture to enable near DRAM memory computing using HBM2 stacking [26]. While Aquabolt-XL houses comparable DRAM memory to a GPU counterpart, it yields limited performance benefits as the logic is much slower and affects the row access latency by up to 2× compared to the standard GPU counterpart [27]. This leads to suboptimal performance exacerbated due to frequent rewrites between the PIM chiplets and the DRAM banks. Disintegrating memory and compute through chiplet-based systems frees up area on the interposer to enable considerable performance benefits. It also enables heterogeneous integration (in terms of technology node, process variation, hierarchical routers).

To summarize, *previous work primarily focuses on accelerating the attention kernel only*. Moreover, none of the prior work considers the role of heterogeneous chiplets along with the associated dataflow to accelerate end-to-end Transformer models. In this paper, we fill this critically important gap in the state-of-art by proposing design principles of a heterogeneous chiplet-based 2.5D manycore architecture to accelerate end-to-end Transformer models.

## 3 TRANSFORMER COMPUTATION KERNELS

This section presents the distinct features of a Transformer model and explains the computational and communication characteristics during the inference process.

### 3.1 Computational Kernels within Transformers

A Transformer primarily comprises an encoder and a decoder stack with a similar structure. The computational structure is identical in Transformer models with varying numbers of encoders/decoder blocks. Fig. 1 show the encoder-decoder structure with their internal computational blocks: it takes an input sequence of length $N$ (e.g., natural text), then it produces an output sequence (e.g., natural text). The encoder has a stack of $k$ identical blocks. Each block consists of two major functional modules: multi-head self-attention and feed-forward (FF). Following these two functional modules, there is a residual block to add the input and the output and to perform the layer norm operation. The multi-head attention layer receives data from the input embedding or previous encoder block. The decoder stack also consists of $k$ identical blocks with an extra cross-attention layer to connect with the output from the last encoder stack.

A Transformer initiates the computation by loading the word embedding $H_{emb}$ and positional encodings $P_{enc}$. Tokenization is the matrix-vector multiplication (MVM) process, where an embedding layer first processes inputs (e.g., words in a sentence) to obtain a learned vector representation of each word in the input sequence. This is a one-time process for an entire Transformer model (encoder-decoder stack).

$$H = H_{emb} + P_{enc}(H_{emb}) \tag{1}$$

After the tokenizing step (a linear MVM operation), each token is represented by a vector of length $d_{model}$, where $d_{model}$ depends on the hidden dimension of the Transformer (e.g., $d_{model}$ = 128 for BERT-Tiny and $d_{model}$ = 768 for BERT-Base)

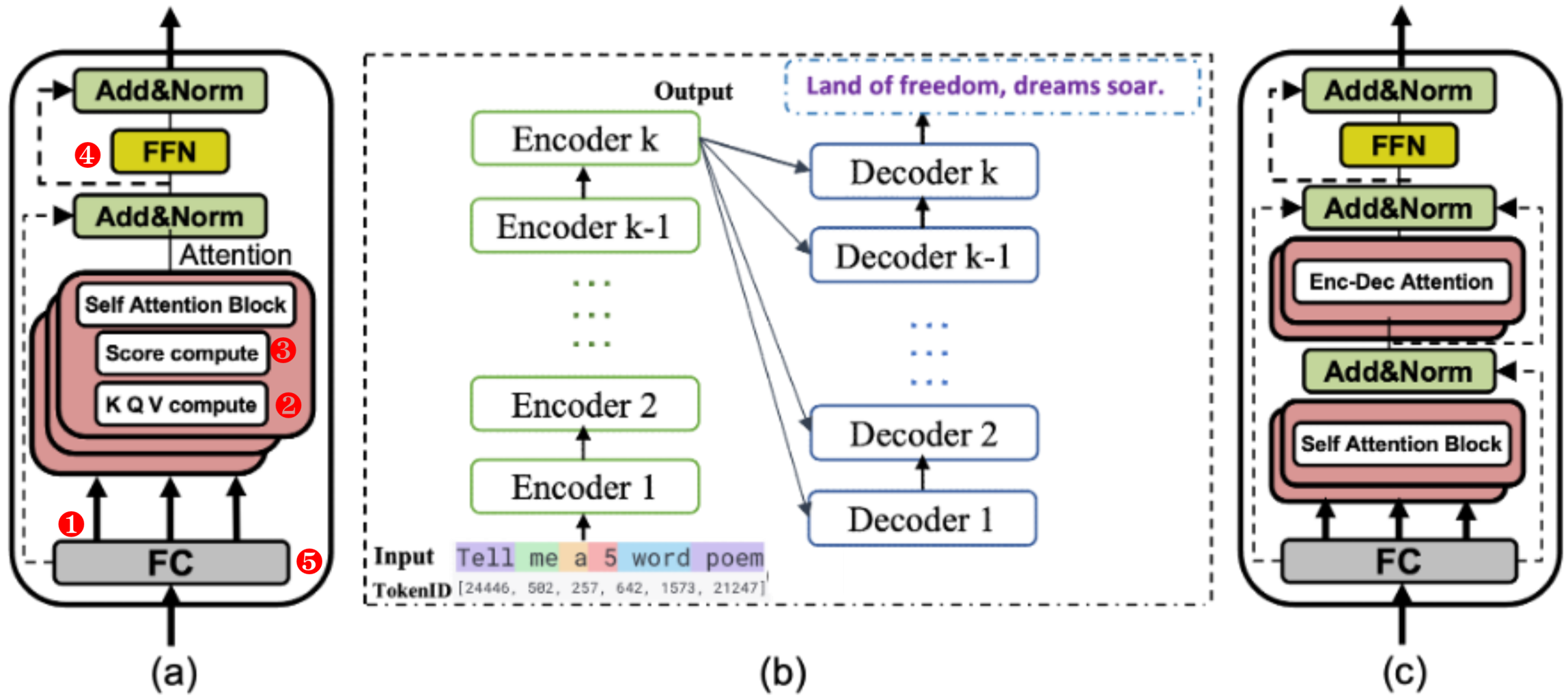

Fig. 1: Overview of the Transformer model; (a) Encoder structure and inherent computational kernels; (b) overall dataflow structure (c) Decoder structure with their inherent computational kernels.

[21]. The corresponding MVM operation is weight stationary and can be effectively computed using processing-in-memory (PIM)-based compute units such as ReRAM chiplets. Next, the weight matrices for the multi-head attention operations are loaded from the memory to the compute units (the structure of each compute unit corresponding to the kernel is explained later). Consistent with the prior work, for $h$ multi heads (each represented with a 16-bit floating point), $h$ separate weight matrices must be loaded to compute $Q_i, K_i, V_i$ shown in (3):

$$\textbf{load } W_i^Q, W_i^K, \text{and } W_i^V \in \mathbb{R}^{\frac{d_{model} X d_{model}}{h}} \tag{2}$$

$$Q_i, K_i, V_i = HW_i^Q, HW_i^K, HW_i^V \ \ \forall\, i \in N \tag{3}$$

Within each self-attention head, the generated tokens are multiplied by distinct weight matrices (distinct for $h$ heads) to obtain the Query $Q$, Key $K$, and Value $V$ matrices (3). For each sequence, the attention generates a representation of the sequence encoded with weighted information from all (or a subset in Masked-Attention) other tokens. This computation depends on the size of the $W_Q$, $W_K$, and $W_V$ matrices. For every self-attention layer within a Transformer encoder, all the compute units must be rewritten before the MVM operation is performed since the *inputs change dynamically for each token* and must be stored in internal registers or on-chip buffers. Traditionally, crossbar-based PIM platforms are employed for the MVM operations involved in DNN workloads [10]. In this case, however, such intermediate matrices will require substantial storage capacity or frequent updates. Hence, traditional nonvolatile memory (NVM)-based PIM architectures are unsuitable here due to their limited write endurance [28]. For example, for BERT-Base and Bert-Tiny, intermediate matrices take up to 8.98× and 2.06× of original weight matrix storage, respectively. The total memory requirements for the two models are 52.8MB and 3.4GB, respectively, when only the weights and positional embeddings are stored [21]. This storage will grow even more for bigger models, which cannot be stored by using more resources and by remaining within the reticle limit of the 2.5D system with an acceptable yield. Hence, the only possibility is to update these intermediate matrices continuously. Consequently, the memory rewrite easily crosses the acceptable endurance limit of the PIM devices [28]. Hence, to implement the attention computation, instead of PIM chiplets, we use the streaming multiprocessors (SMs) along with the associated memory controllers (MCs) to access the weights $W$ from the DRAM chiplet. All the $K, Q, V$ computations are executed parallelly, leading to a many-to-few traffic scenario (multiple SMs and a few MCs).

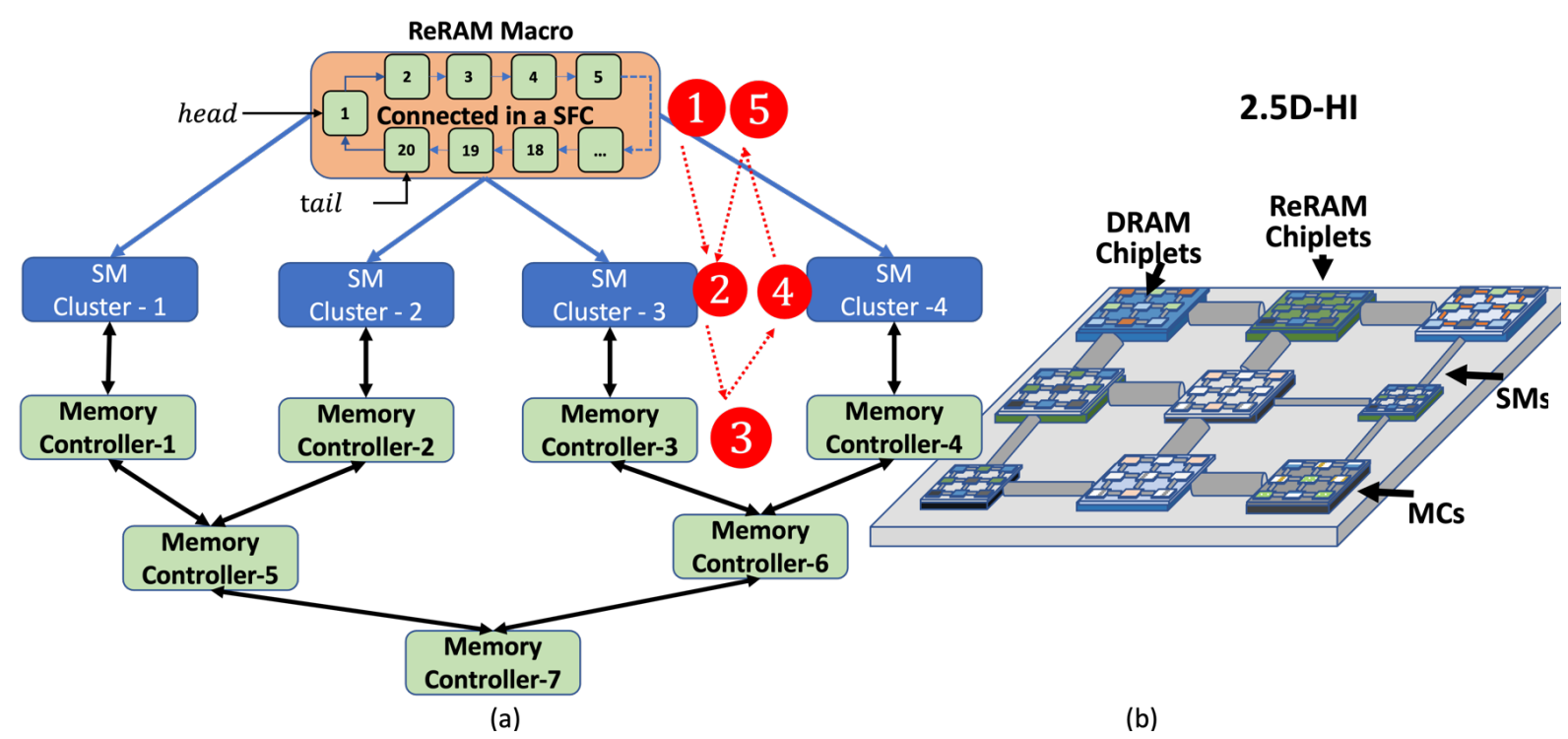

Fig. 2: Overview of the (a) proposed heterogeneous architecture dataflow. ❶-❺ describe the dataflow within each computation kernel corresponding to Fig. 1; (b) 2.5D heterogeneous integration for accelerating Transformer model. It should be noted that the figure is an illustration and not drawn to scale.

Next, the outputs, Query, and Key matrices are multiplied and normalized with a hidden dimension $d_{model}$ *by* a Softmax layer to generate the intermediate attention probabilities (4) & (5).

$$\alpha_i = Q_i K_i \tag{4}$$

$$Score_i = softmax\left(\frac{\alpha_i}{\sqrt{d_{model}}}\right) \tag{5}$$

The intermediate attention probabilities are then multiplied with the Value matrix to obtain the final output (6).

$$P_i = Score_i V_i \tag{6}$$

Multiple attention heads are processed in parallel, and the resultant matrices are concatenated and multiplied with the weight matrix $W_i^O$, which maps the attention probabilities to output scores before advancing the output to the FF layers where $N$ is the input-sequence length (given by (7)).

$$H^{MHA} = concat\left(P_i W_i^O\right) \forall\, i \in N \tag{7}$$

Once the score is computed within the multi-head self-attention block, we add the input to the output of the multi-head attention and normalize the resultant matrix. This step is the layer-norm operation employed to reduce covariance shifts within the fully connected (FC) layer [21]. GeLU is the commonly used activation function in Transformers [2] [29] [30]. For modern large language models (LLMs), FC layers dominate the runtime of the decoder block since the model dimension $d_{model}$ is much larger than the input sequence length $N$ and the context length $l_{context}$, thus $O(Nd_{model}{}^2) >> O(l_{context}^2 d_{model})$. For example, more than 99% of MVM operations in GPT-3 are spent on executing the FC layers. Since the FC layers are not frequently updated, they are static and have significant weight overlap in each encoder-decoder stack. Therefore, we utilize ReRAM chiplets for the FF [15].

To summarize, we use a combination of SM-MC-DRAM and ReRAM chiplets placed on a 2.5D interposer to execute the end-to-end Transformer models. Due to this heterogeneity, analyzing the inter-chiplet traffic patterns is imperative to design the overall NoI architecture.

### 3.2 Communication between Computation Kernels

Fig. 2(a) shows the overall dataflow in the heterogeneous chiplet platform. Various components of the NoI traffic are:

***Input Embedding**:* The input embedding provides each token with an input representation. The tokenization is a series of MVM on the input sequence where the overall computation is sequentially distributed among multiple ReRAM chiplets. Data always flows from the $i^{th}$ to the $(i+1)^{th}$ chiplet. Hence, contiguity should be maintained on the physical NoI layer, to the extent possible, between any two consecutive chiplets to reduce the communication overhead. Since existing NoI architectures are primarily based on standard multi-hop regular topologies such as a mesh or a torus, it may not always be possible to find contiguously placed chiplets available to map successive parts of the input embedding. If two consecutive layers are mapped far apart, it will lead to long-range multi-hop communication through the NoI. This, in turn, will degrade the performance and energy efficiency of the NoI and hence the overall system. Therefore, we connect the ReRAM chiplets using space-filling curves (SFCs). This dataflow-awareness within computational kernels has been previously exploited for designing a NoI called Floret for large-scale CNN inferencing using SFC [9] [31]. SFCs are a specialized class of algorithmic mapping techniques that have found significant application in locality-preserving data structure for numerous scientific applications that involve spatial and range queries [32] [33]. More specifically, an SFC maps a multi-dimensional point cloud onto a single dimension; therefore, each SFC represents a linear ordering of the input set of points. Numerous types of SFCs have been defined over the decades, including simple schemes such as row/column major curves to more sophisticated curves such as the Hilbert curve, Morton or Z-curve, or onion curve. For a review of classical SFCs, please refer to [33] [34] [35].

This operation is referenced as ❶ in the Fig. 2(a). Our approach connects the ReRAM chiplets (in the order of data flow) along the contiguous path formed by the SFC. The SFC stitches multiple chiplets contiguously and has a head and a tail that connect the chiplets linearly (Fig. 2 (a)).

***K Q V Computation**:* We implement this step, which is the starting part of the self-attention mechanism using SM, MC, and DRAM chiplets. After computing each input token generated in Step 1, $W_k, W_{Q,}, W_V$ need to be loaded. We use the FlashAttention dataflow to partition the matrices onto the SMs [36]. This operation can be effectively performed using DRAM (HBM2) with multiple SM chiplets for each MC partition. The data transmission happens in two steps. First, for each head $h$, $W_k, W_{Q,}, W_V$ are loaded in the SMs through the MCs. Then, the corresponding $K, Q, V$ is computed for each input token. This gives rise to SM to MC data exchange. Due to many SMs and a limited number of MCs, this results in a many-to-few data flow. These steps are shown as ❷ & ❸ in Fig. 2(a). The placement of the DRAM chiplet also contributes to the overall latency and hence the placement is imperative with respect to other computing chiplets. We consider SM, MC, and DRAM chiplets are all distributed over the 2.5D interposer. We call it 2.5D heterogeneous integration (2.5D-HI), shown in Fig. 2(b).

***Score Computation**:* This is the final step of the self-attention to multiply the attention score matrix $S$ after Softmax by the $V$ matrix before the FF layer (shown as ❹ in Fig. 2(a)). The placement of SM and MC chiplets is critical for the latency and energy consumption.

***Feed Forward Layer**:* The FF network consists of two consecutive FC layers, which are large static hidden layers. Like DNN models, the fully connected layers have fixed sizes and sparse weight updates compared to encoder outputs. The FC layers have dataflow from layer $L_i$ to layer $L_{i+1}$ connection. The activations and corresponding weights of the adjacent layers must be mapped in a contiguous manner. Hence, there is a need to maintain contiguity on the physical NoI layer, to the extent possible, between the neural layers to reduce communication overhead. This is identical to the dataflow pattern observed in the case of Input Embedding generation. Hence, like ❶, an SFC is employed to maintain contiguity among communicating layers mapped on the ReRAM chiplets. The ReRAM macro refers to a set of ReRAM chiplets employed to execute computations in the feed-forward layers contiguously, thereby reducing communication overhead within a Network on Interconnect (NoI). This part of the execution is shown as ❺ in Fig. 2(a). We repeat the above-described dataflow loop starting from ❷ for the next encoder as Input Embedding is a one-time process in the overall execution flow. It should be noted that the same ReRAM macro is used for ❶ and ❺ as they execute on different time stamps. This macro of ReRAM chiplets maintains data flow between consecutive layers efficiently, leveraging SFCs to preserve locality and enhance the performance and energy efficiency of the overall system. Hence, we consider all the ReRAM chiplets together as a *ReRAM macro.* In the NoI design, we optimize the location of the ReRAM macro with respect to SM, MC, and DRAM chiplets.

The original transformer model with encoder-decoder blocks is designed for translation-class of natural-language processing (NLP) tasks [37]. However, the transformer architectures evolve continuously with structural variations tailored to other NLP tasks. Several transformer architectures have recently been proposed, including transformers composed of decoder-only (GPT-class of transformer models) or encoder-only (BERT-class of transformer models) blocks. This effectively divides the model complexity in half, reducing the computational requirements.

Additionally, numerous architectural refinements have also emerged within the encoder/decoder block. One such advancement is the introduction of Multi-Query Attention (MQA). MQA is specifically designed to enhance the efficiency of attention mechanisms [38] [30]. MQA, unlike the standard multi-head attention (MHA), utilizes the same K and V values across heads while assigning distinct Q vectors to each head. Fig. 3 shows the difference between MHA and MQA. MQA shares a single key and value head across all query heads. The amount of computation performed by MQA is equivalent to the standard MHA. However, the key difference is the reduced amount of data exchange from memory to computing chiplets.

Another variation is the parallel attention framework, which executes the model with both the attention and FF layers operating concurrently [30]. The model modifies the computational kernel in each transformer block and employs a parallel formulation as opposed to the conventional "serialized" formulation. The standard formulation, in this context, can be expressed as:

$$Y = x + MLP\Big(LayerNorm\big(x + Attention(LayerNorm(x))\big)\Big) \quad (8)$$

In contrast, the parallel formulation can be described as:

$$Y = x + MLP\big(LayerNorm(x)\big) + Attention\big(LayerNorm(x)\big) \quad (9)$$

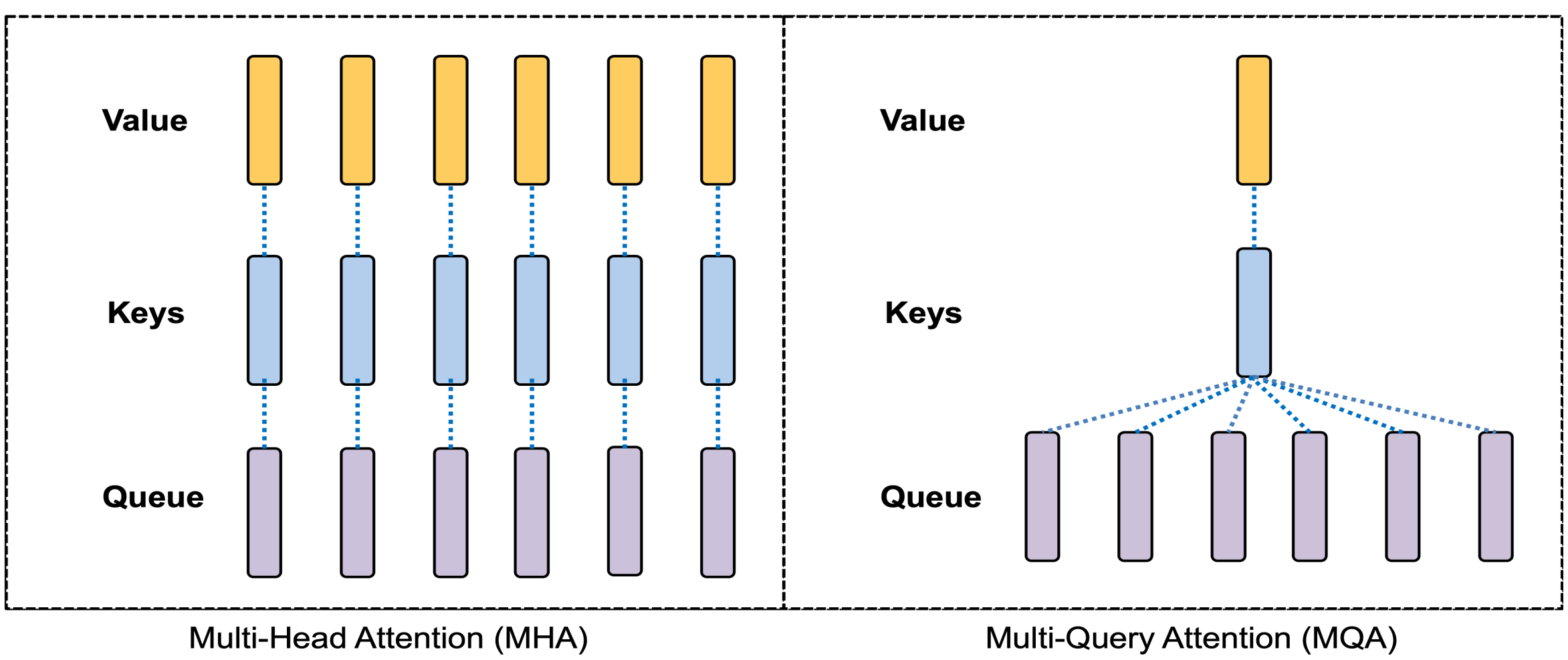

Fig. 3: High-level illustration of the difference between multi-head attention (MHA) and multi-query attention (MQA).

The parallel formulation (given in (9)) facilitates a pipelined execution of attention and feed-forward computations, thereby achieving higher speedup. We consider all these variations (encoder/decoder-only, MQA, and parallel attention) and propose a model-agnostic NOI tailored for different classes of transformer models.

### 3.3 NoI Design Optimization

Considering the traffic patterns mentioned above, we need to place the DRAM, the ReRAM Macro, SM, and MC chiplets on the interposer to maximize throughput and minimize the data exchange latency to accelerate the Transformer model inferencing. Suppose each candidate design ($\lambda$) in the NoI design space corresponds to a specific placement of the chiplets along with the routers and inter-router links. Our goal is to achieve high throughput under the given application traffic pattern. Minimizing the mean $\mu(\lambda)$ and standard deviation $\sigma(\lambda)$ of the traffic distribution leads to an overall higher throughput of the candidate NoI designs [10] [39].

**MOO Formulation:** The design of heterogeneous NoI architecture can be formulated as a multi-objective optimization (MOO) problem. We can represent the MOO formulations as follows:

$$\lambda^* = MOO\big(objective = \mu(\lambda), \sigma(\lambda)\big) \tag{10}$$

where $\lambda^*$ is the set of Pareto optimal designs. Here, we aim to find the optimal placement of chiplets and associated routers and links such that the design requirements for all the elements are satisfied. Each candidate NoI design must be evaluated efficiently to help guide our search for the best NoI architecture. Below, we describe the critical elements of our MOO formulation.

1) **Design Variables:** There are two types of design variables for NoI-based chiplet design $\lambda = (\lambda_c, \lambda_l)$, where $\lambda_c$ corresponds to a candidate placement of chiplets and $\lambda_l$ corresponds to a candidate placement of communication links between chiplets.
2) **Constraints:** We have two constraints to define a feasible solution space. First, the NoI should enable data transfer between any two pair of chiplets, i.e., it should connect all chiplets with no islands. Second, the number of links used to design NoI should not exceed that in a 2-D mesh. However, with an efficient NoI, we can reduce the number of links compared to a mesh NoI [10].

   **Objectives:** Our objective is to maximize the system throughput. Specifically, we consider optimizing two objectives, namely, the mean $\mu(\lambda)$ and standard deviation $\sigma(\lambda)$ of the traffic utilization $u_k$ on each communication link k at time t, determined by the NoI frequency. Note that the $F_{ij}$, the amount of traffic between chiplets $i$ and $j$, respectively, can be obtained by profiling the application workload, i.e., inference of Transformer models.

These parameters are defined as follows:

$$u_k = \sum_{i=1}^{L}\sum_{j=1}^{L} F_{ij}(t) \cdot q_{ijk} \tag{11}$$

$$\mu(\lambda, t) = \frac{1}{P} \times \sum_{k=1}^{P} u_k \tag{12}$$

$$\sigma(\lambda, t) = \sqrt{\frac{1}{P}\sum_{k=1}^{P}\big(u_k - \mu(\lambda, t)\big)^2} \tag{13}$$

where $L$ represents the number of routers and $P$ represents the number of links in the system, as mentioned earlier, $q_{ij}$ is a Boolean representation if link $k$ is being used between router $i$ and router $j$.

$$q_{ijk} = \begin{cases} 1, & \text{if chiplet i, j communicate across link k.} \\ 0, & \text{otherwise} \end{cases}$$

To compute the average throughput across all timestamps, we take a time average of the mean $\mu(\lambda)$ and standard deviation $\sigma(\lambda)$ as represented in (14) and (15):

$$\mu(\lambda) = avg\ \mu(\lambda, t) \tag{14}$$

$$\sigma(\lambda) = avg\ \sigma(\lambda, t) \tag{15}$$

Once we solve the MOO problem to obtain the Pareto optimal NoI designs with proper placement of chiplets and inter-chiplet links ($\lambda^*$) among all possible system configurations, we perform cycle-accurate simulations for each design in $\lambda^*$ to find the design with the lowest EDP and lower latency than a mesh NoI simultaneously.

**MOO Solver:** The critical challenge in solving the MOO problem for our chiplet-based design is the vast combinatorial design space. We need to consider designs with fewer than or equal to the number of links in a 2D mesh NoI for energy efficiency through smaller routers. Existing MOO methods consider a significantly smaller size of the design space and are slow in converging to Pareto optimal design set. Simulated annealing-based AMOSA algorithm has been applied for the problem of heterogeneous NoC design in the past [40]. AMOSA is shown to be superior to genetic algorithm (GA), or branch and bound algorithms [41] [42]. However, since AMOSA is based on simulated annealing, it needs to be annealed slowly to ensure a good solution, which does not scale well with our expected search space [39]. Prior work has shown that we can significantly improve the accuracy and speed of design optimization by utilizing knowledge gained from the past explored designs. The key idea is to extract relevant knowledge from previously explored designs to intelligently guide the search to more promising parts of the design space. In this work, we adopt the MOO-STAGE, a MOO algorithm that belongs to this class of data-driven methods, for the problem of NoI design, noting that any other MOO solver can be used. MOO-STAGE is an iterative algorithm which learns an evaluation function to select good starting states to guide local search procedures such as greedy search toward better local optima [10] [39]. Each iteration consists of the following three steps: First, we use the current evaluation function to select a good starting state for the local search procedure. Second, we perform a local search from this selected starting state until we reach a local optimum and compute the quality of the corresponding Pareto set in terms of Pareto-hyper volume (PHV) as a metric. Third, we update the evaluation function based on the training data from the local search runs. As the evaluation function improves with the training data over iterations, it prunes the design space and directs the search toward promising regions [39].

***Learning Evaluation Function:*** The goal of the evaluation function is to estimate the quality of the Pareto set obtained by performing a local search from a given starting design. Given a learned evaluation function, we perform a local search guided by this evaluation function until we reach a local optimum to select the starting state (Meta search) for the actual local search process (Base search). If the evaluation function is accurate, it will avoid bad starting states and select promising starting designs to improve the efficiency and accuracy of design optimization. Otherwise, we update the evaluation function to minimize errors. Each training instance in MOO-STAGE corresponds to the past local search run: sequence of designs $(d_1, d_2, \ldots., d_T)$ forming the local search trajectory and the quality of the resulting Pareto set from the local optima measured in terms of PHV metric. We create one regression training example for each design $d_i$ on the local search trajectory: $d_i$ is the input and PHV of the local optima is the output. We give the aggregate set of regression examples to the random forest algorithm to create/update the evaluation function. We use random forest as it was shown to be a fast and accurate learner to create effective evaluation functions [10]. Fig. 4 shows the pareto optimal curve for different design variables considered in this work for accelerating transformer models.

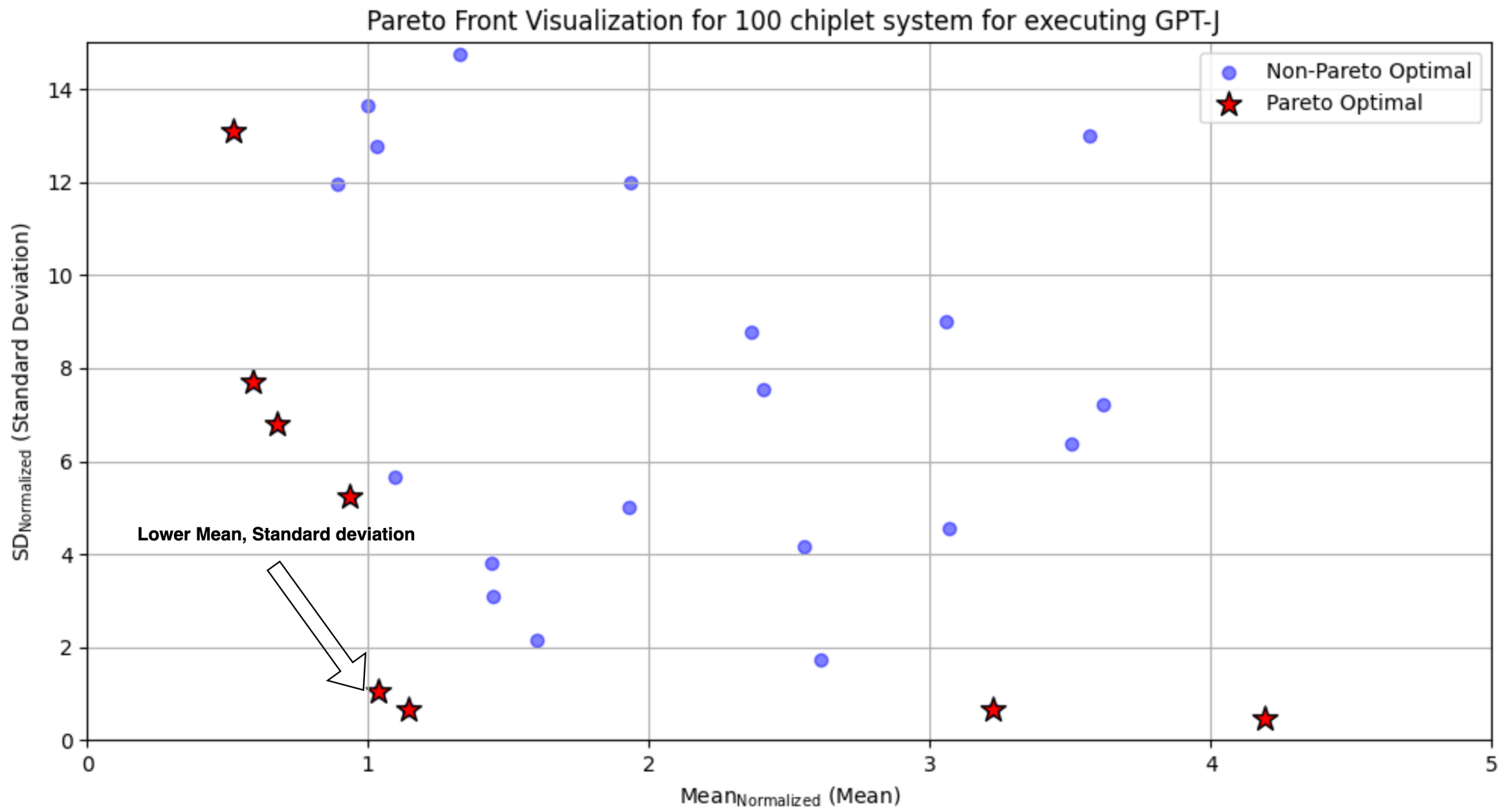

Fig. 4: Pareto optimal points when evaluating different architectural design choices. Both the x- and y- axis are normalized with respect to a standard 2D-Mesh connections. A lower mean and standard deviation is preferred for higher performance and energy efficiency of the interconnection network.

## 4 EXPERIMENTAL RESULTS

In this section, we first introduce the experimental setup to evaluate the performance of the proposed HI architecture. Next, we describe the transformer model considered in our evaluation. Subsequently, we present a comparative performance evaluation of the proposed 2.5D-HI architecture with respect to state-of-the-art baselines.

### 4.1 Experimental Setup:

#### 4.1.1 System Specification and Evaluation Setup:

Our performance evaluation considers three system sizes of 36, 64, and 100 chiplets. Fig. 5 shows the design specification of each computational platform employed in the 2.5D chiplet-based systems. The chiplet system employs SM, MC, DRAM and ReRAM chiplets. We maintain SM-MC configurations consistent with the Volta architecture [27] [43]. Each MC is connected to a DRAM chiplet. To illustrate, in the case of a 100 chiplet system, we divide the 2.5D-HI system into 64 SMs, 8 MCs along with 8 DRAM chiplets corresponding to 8 partitions of the HBM2 memory, and 20 ReRAM chiplets connected as a ReRAM macro. Note that this is an example system size, and the proposed design methodology and evaluation scheme are valid for other system configurations.

The ReRAM chiplets perform the Feedforward (FF) network computation where the weights are spatially partitioned across the ReRAM chiplets. Our approach connects the ReRAM chiplets (in the order of data flow) along the contiguous path formed by the SFC. Subsequently, a pipelined implementation is utilized across the FF network layers where the output from MHA computation serves as input to the pipeline. The number of ReRAM crossbars (chiplets) necessary to map the FF weights vary based on the transformer model size. Thus, we utilize a strategy of weight duplication in scenarios where the model's weight parameters can be mapped on a subset of the available ReRAM chiplets. The duplication of weights is a common strategy to enhance computation throughput on ReRAM-based architectures [44]. It creates copies of the layer weights and parallelizes the computation across multiple ReRAM crossbars [44]. This helps prevent any underutilization of ReRAM chiplets during FF computation with varying model sizes. Table 2 presents the allocation for the three-system sizes considered in this work.

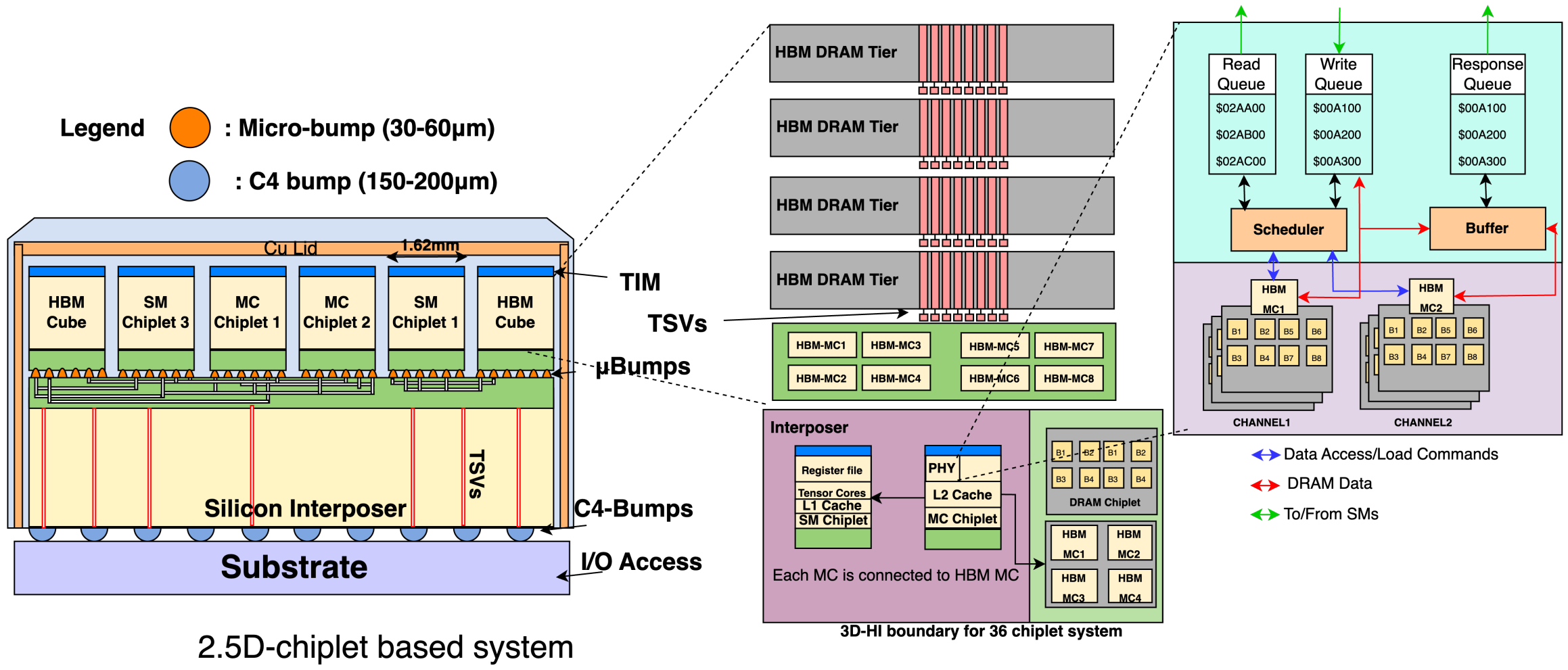

Fig. 5: Illustration of 2.5D-HI integration for accelerating Transformer model.

SMs are based on the NVIDIA Volta architecture with 10 Tensor cores along with the scratchpad memory for arithmetic operations. The group of collocated SMs (SM cluster in Fig. 2 (a)) are associated with a particular MC. Following prior work, we consider each ReRAM chiplet to consist of 16 ReRAM tiles and peripheral circuits such as accumulator, buffer, activation, and pooling units [10]. Each tile consists of multiple processing elements (PEs) comprising 128×128 ReRAM crossbar arrays. Within each tile 96 PEs are connected with an H-Tree-based point-to-point network [10] [45].

**DRAM Microarchitecture:** The DRAM used in the proposed chiplet-based architecture is an industry standard High Bandwidth Memory (HBM2), which is a multi-layered DRAM technology specifically designed to offer high bandwidth for data-intensive applications such as the transformer acceleration considered in this paper. We do not integrate any custom computing hardware within HBM in our proposed architecture. The HBM2 uses Through-Silicon Vias (TSVs) to connect multiple 3-D stacked DRAM tiers on top of a base logic die. Each stack has two independent HBM channels where the operation of channels, including those within the same core die, is independent via private data and address/control signals and buses. Each channel utilizes dedicated 128-bit non-shared TSV connections for data exchange. Overall, the memory bandwidth provided by the HBM is determined by the number of DRAM stacks, with each stack providing a 256-bit wide interface via two channels. Further, each channel has an associated HBM memory controller (HBM-MC1, HBM-

Table 2: Resource allocation among SM, MC, and ReRAM chiplets for different system sizes

| System Size (No. of Chiplets) | SM | MC | DRAM | ReRAM |
|---|---|---|---|---|
| 36 | 20 | 4 | 4 | 8 |
| 64 | 36 | 6 | 6 | 16 |
| 100 | 64 | 8 | 8 | 20 |

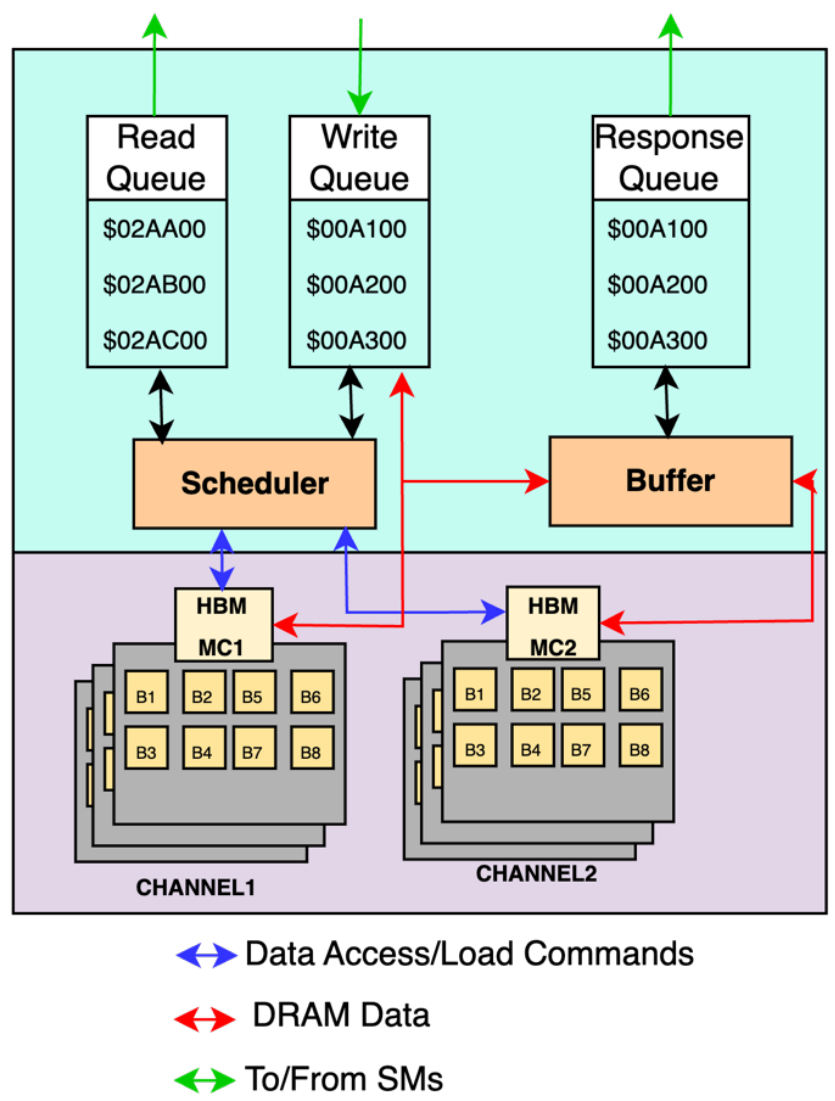

Fig. 6: Illustration of communication protocol between the DRAM chiplets and the (memory controller) MC chiplets.

MC2, and so on), as shown in Fig. 5 in the logic die. The function of the HBM-MC is to enable data exchange with the MC chiplet. The communication protocol between the HBM-MC and the HBM channels is facilitated through a First-In-First-Out (FIFO) based interface within the scheduler block, as shown in Fig. 6. This interface is partitioned into distinct FIFOs, allocated for logical address, write, and read data.

The HBM-MC performs address mapping and reads/writes data when a read/write request is issued. We connect each HBM-MC to the scheduler within the MC chiplet in a point-to-point manner, as shown in Fig. 6. Each MC chiplet is equipped with a dedicated physical layer IP interface (PHY) responsible for providing a low-level physical interface to the associated HBM channel through the HBM-MC. The PHY utilizes an industry standard interface protocol (DFI) between the HBM-MC and the MC chiplet [46]. The role of the PHY encompasses the generation of all handshake signals with precise timing requirements necessary for interfacing.

Depending on the system size, we utilize three distinct HBM2 configurations in this work. Specifically, we employ: 2-tier, 3-tier, and 4-tier DRAM stacks, correspondingly tailored to accommodate the memory demands of systems comprising of 36, 64, and 100 chiplets, respectively. It should be noted that HBM2 can support up to 8 DRAM stacks.

Here, MC and DRAM chiplets are allocated in a 1:1 ratio. Each DRAM chiplet corresponds to an HBM channel. This uniform allocation is necessary since the interface communication protocol between HBM-MC and MC chiplet requires point-to-point communication to facilitate data transfer to and from the HBM.

The utilization of SM, MC, and ReRAM chiplets is independent of the transformer model parameters. Specifically, SM chiplets execute each computation in multiple parallel threads with instruction and data accessed through the MC. The number of threads for each MHA computation is orders of magnitude higher than the available SMs. Consequently, this prevents any underutilization scenario for SM-MC chiplets based on the model size for the MHA computation.

***Evaluation Tool Flow:*** The weights for each computational kernel are loaded through the DRAM chiplets, stacked on top of the interposer shown in Fig. 5. We obtain the workload traces using Nvidia-smi with Nvidia A40 GPU to evaluate the performance and energy efficiency of the system. The software stack for running the Transformer model to obtain the traffic pattern is built on top of the CuDNN 12.2 with PyTorch, consistent with prior work [20]. We develop a tool flow

Table 1: Design specifications for computational platforms employed in chiplet-based system.

| | |
|---|---|
| **ReRAM Chiplet: 16-tiles per chiplet** [66] | |
| ReRAM Tile | 96-ADCs (8-bits), 12×128×8DACs (1-bit), 96 crossbars, 128×128 crossbar size, 2-bit/cell resolution, 0.34 W, 0.37 $mm^2$, Tech node – 32 nm [66] |
| **SM-MC Chiplets:** [43] | |
| SM Chiplet | Volta architecture, 10 Tensor core, 64 KB SM register file, 96 KB L1 cache, 1530 MHz |
| MC Chiplet | L2 cache – 512 KB, 3.2 mm2, Tech node – 12 nm |
| **DRAM Chiplets:** [26] | |
| DRAM | 1-4 tier; 2 channel per tier; 16 banks per channel; 2GB per channel; Tech node – 12 nm |
| **Interposer Parameters – 65nm** [7] | |
| $\mu$Bump dimension - 25$\mu$mX10$\mu$mX50$\mu$m; C4 Bump dimension - 250$\mu$mX70$\mu$m X600$\mu$m; Wire Thickness - 1$\mu$m; Wire width – 1.5$\mu$m; Link frequency - 0.6ns/mm; Link length – 1.449mm | |

(shown in Fig. 7) integrating NeuroSim, BookSim2, *AccelWatch, Nvdia-smi,* and VAMPIRE to evaluate the performance and power consumption of each architectural choices under consideration [47] [48] [49] [50]. The red block represents the inputs to the evaluation framework, while the purple blocks are the outputs. The input to the evaluation framework is the LLM model, the location of chiplets, their interconnection network, and link configuration. NeuroSim outputs the expected compute power and latency for the ReRAM chiplets. BookSim2 simulator outputs communication power and NoI latency for the entire system; VAMPIRE is used to estimate the DRAM power and execution time. We use a modified version of the VAMPIRE simulator to compute access times (employing RAMULATOR) and expected energy consumption at 500MHz following the configuration illustrated in Table 1. The inter-chiplet traffic flowing through the NoI is generated by the activations exchanged between the various chiplets. These include data exchange between the SM-MC chiplets, among the ReRAM chiplets, as well as between ReRAM macro and MC chiplets. All the NoI topologies, considering the detailed model based on Nvidia GRS parameters, are then simulated using the BookSim2 simulator [11] [51]. We employ the tiling technique for matrix multiplication in SMs, where input data blocks are loaded from DRAM to MC. Tiling is a standard program transformation technique that localizes the memory locations accessed among threads and the timing of their accesses. *AccelWatch* considers the tiling strategy used in the GPU architecture, including the tile size, shape, and overlap, to estimate the power consumption of the GPU. This allows for more accurate power modeling, especially in scenarios where tiling has a significant impact on power consumption, such as in the case of Transformer models. Finally, a thermal simulator (Hotspot 6.0) is used to model the thermal profile [52]. It should be noted that other thermal simulators like MFIT, 3D-ICE, or PACT may also be used for similar accuracy [53] [54] [55]. For ReRAM chiplets, we use a modified

Table 3: Transformer Model and Parameters

| Model | Transformer Architecture | $d_{model}$ | Layers | Heads | Parameters (In millions) |
|---|---|---|---|---|---|
| **BERT-Base** | Encoder Only | 768 | 12 | 12 | 110 |
| **BERT-Large** | Encoder Only | 1024 | 24 | 16 | 340 |
| **BART-Base** | Encoder-Decoder | 768 | 12 | 12 | 140 |
| **BART-Large** | Encoder-Decoder | 1024 | 12 | 16 | 400 |
| **GPT-J** | Decoder Only | 4096 | 28 | 16 | 6700 |
| **Llama2-7B** | Decoder Only | 4096 | 32 | 32 | 7000 |

NeuroSim to implement the MVM operations involved in the Input Embedding and the FF layers in the Transformer model [47]. Further, we use the cycle-accurate BookSim2 simulator to implement the interconnection network to connect the chiplets via different NoI architectures [31] [48]. The inputs to the BookSim2 are the connectivity between NoI routers and the inter-chiplet traffic traces for the Transformer model. The simulator outputs area, latency, and NoI energy consumption. Considering the clock frequency of 1.2GHz, each 1.55 mm link can be traversed in one cycle [7]. The longer links are divided into multiple stages, where each stage is 1.55 mm long. We employ the Nvidia ground-referenced signaling (GRS) parameters for an interconnection network designed using 32nm technology to evaluate the NoI area and power consumption [11].

We compare the performance and energy efficiency of the 2.5D-HI architecture with the recently proposed state-of-the-art HAIMA and TransPIM architectures on various sequence lengths [2] [3]. We redesigned HAIMA and TransPIM on the chiplet-based system as HAIMA_chiplet and TransPIM_chiplet, respectively. In both HAIMA_chiplet and TransPIM_chiplet, we implement the same MOO algorithm as the proposed HI configurations to suitably place the chiplets, routers, and associated links to give a fair optimization as we follow in comparative heterogeneous architectures. We maintain an iso-chiplet configuration across all architectures and system sizes, where the physical area of each chiplet is fixed.

### 4.1.2 Datasets and Transformer Workloads

We consider six transformer models in our experimental evaluation: BERT-Base, BERT-Large, BART-Large, Llama2-7B, and GPT-J [30] [38] [37]. These models represent diverse transformer architectures, including Encoder-Decoder, Encoder/Decoder only, MQA, and parallel MHA-FF architectures. BERT adopts an encoder-only transformer architecture, while BART incorporates both encoder and decoder blocks. Llama2-7B and GPT-J are decoder-only models. Llama2-7B uses MQA instead of MHA for attention computation, whereas GPT-J employs Parallel MHA-FF computation. Table 3 presents the different Transformer models considered in our experimental evaluation and their details, including model dimensions, number of heads, number of layers, and number of parameters.

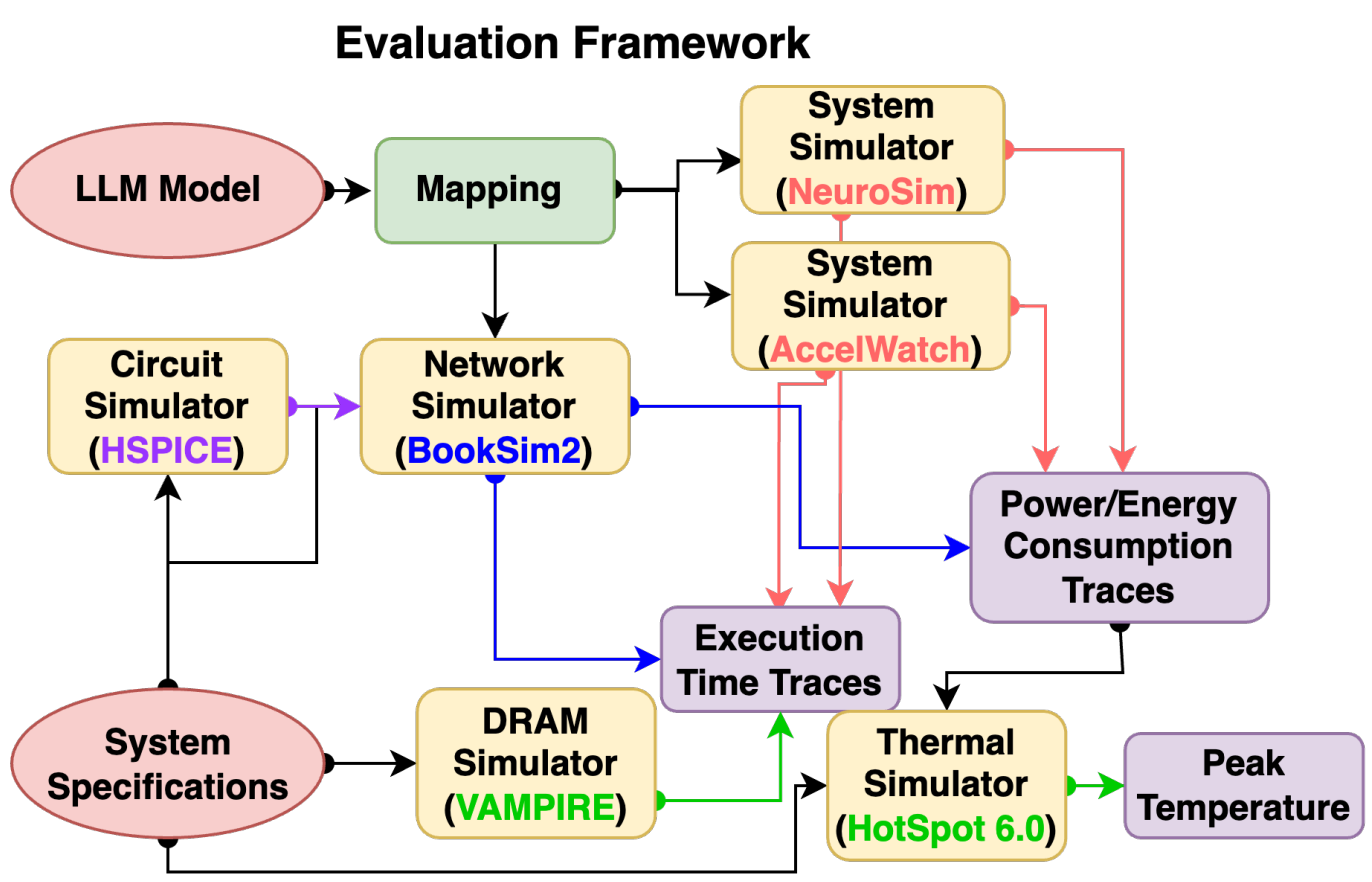

Fig. 7: Illustration of tool flow for evaluating chiplet based heterogeneous integration for accelerating Transformer model.

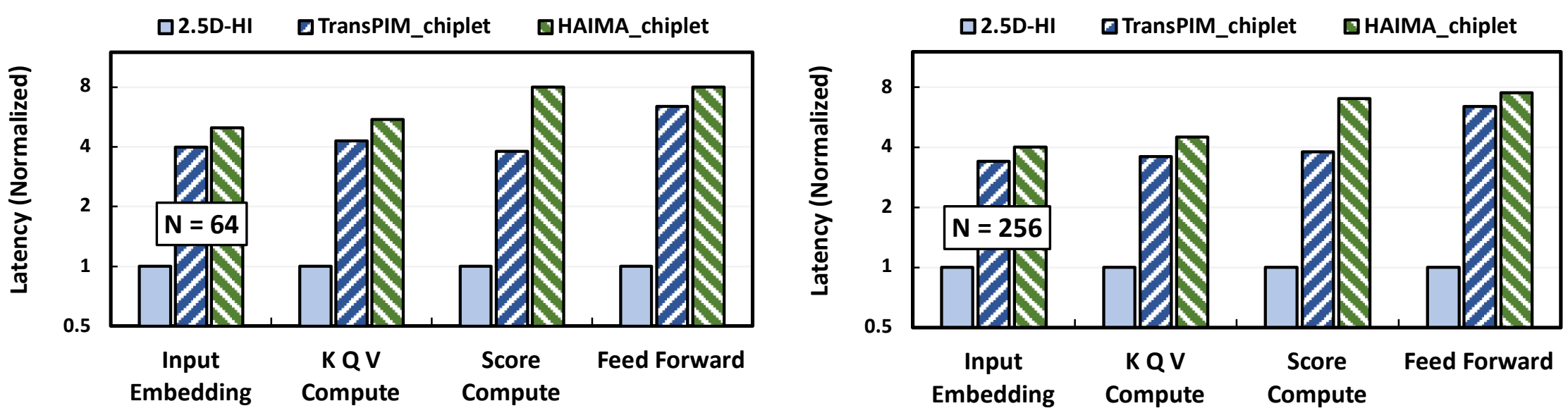

Figs. 8: Latency improvement for each computational kernel of the NoI architecture comparing 2.5D-HI with counterparts for (a) *N*=64 & (b) *N*=256 for 36 chiplet system.

First, we evaluate a 36-chiplet system with BERT-Base. As the system size increases, we progressively employ models with increased parameter counts to illustrate the scalability of the proposed HI architecture. Specifically, we extend our evaluation to a 64-chiplet system with BERT-Large and BART-Large. Finally, we scale up to a 100-chiplet system with Llama2-7Bs and GPT-J, where the total number of parameters are in billions.

### 4.2 Effect of Heterogenous Integration on Performance

We undertake a full system performance evaluation and compare the overall execution time and energy consumption of 2.5D-HI with respect to the TransPIM_chiplet and HAIMA_chiplet for multiple sequence lengths. Following prior work, TransPIM_chiplet consists of multiple auxiliary compute units (ACUs) to enable computing near DRAM chiplets. ACUs perform vector reduction and Softmax function [2]. Other computational kernels are calculated with a bit-serial row-parallel scheme. The token sharding technique is employed as the optimal dataflow strategy. This enables DRAM chiplets to compute the partial attention score matrix independently without communicating with other chiplets. In the case of HAIMA_chiplet, SRAM chiplets compute specific parts of the score computation (equation (5) & (6)). DRAM chiplets are employed for implementing self-attention and FF layers. Following the original HAIMA framework, we use multiple host chiplets to compute the arithmetic operations [3].

Figs. 8(a) & (b) show the latency profiles for different computational kernels for 2.5D-HI, TransPIM_chiplet, and HAIMA_chiplet for two sequence lengths (64 and 256) as examples for 36 chiplet system. We observe that 2.5D-HI has lower latency than both TransPIM_chiplet and HAIMA_chiplet for all the computing kernels. For self-attention (K, Q, V, and Score computation), we see consistent performance gains over TransPIM_chiplet and HAIMA_chiplet. The tensor cores of the SMs in 2.5D-HI enable the performance gain for self-attention. However, the performance gain is maximum for the FF layer due to the efficient computation by ReRAM chiplets. The SMs efficiently accelerate MHA computation, and the ReRAM layer computes the FF layer in parallel. In contrast, both HAIMA and TransPIM have latency overhead in a chiplet-architecture. HAIMA maximizes the throughput by activating multiple banks within the DRAM-PIM parallelly. In 2.5D systems, these banks need to be disintegrated into chiplets, which leads to higher power consumption and frequent data exchange between SRAM and DRAM chiplets and gives rise to multiple contention paths. This leads to higher latency overheads and reduced energy efficiency compared to 2.5D-HI and a simpler ring broadcast-based communication dataflow proposed in TransPIM. Additionally, 2.5D-HI benefits from the fused score and Softmax

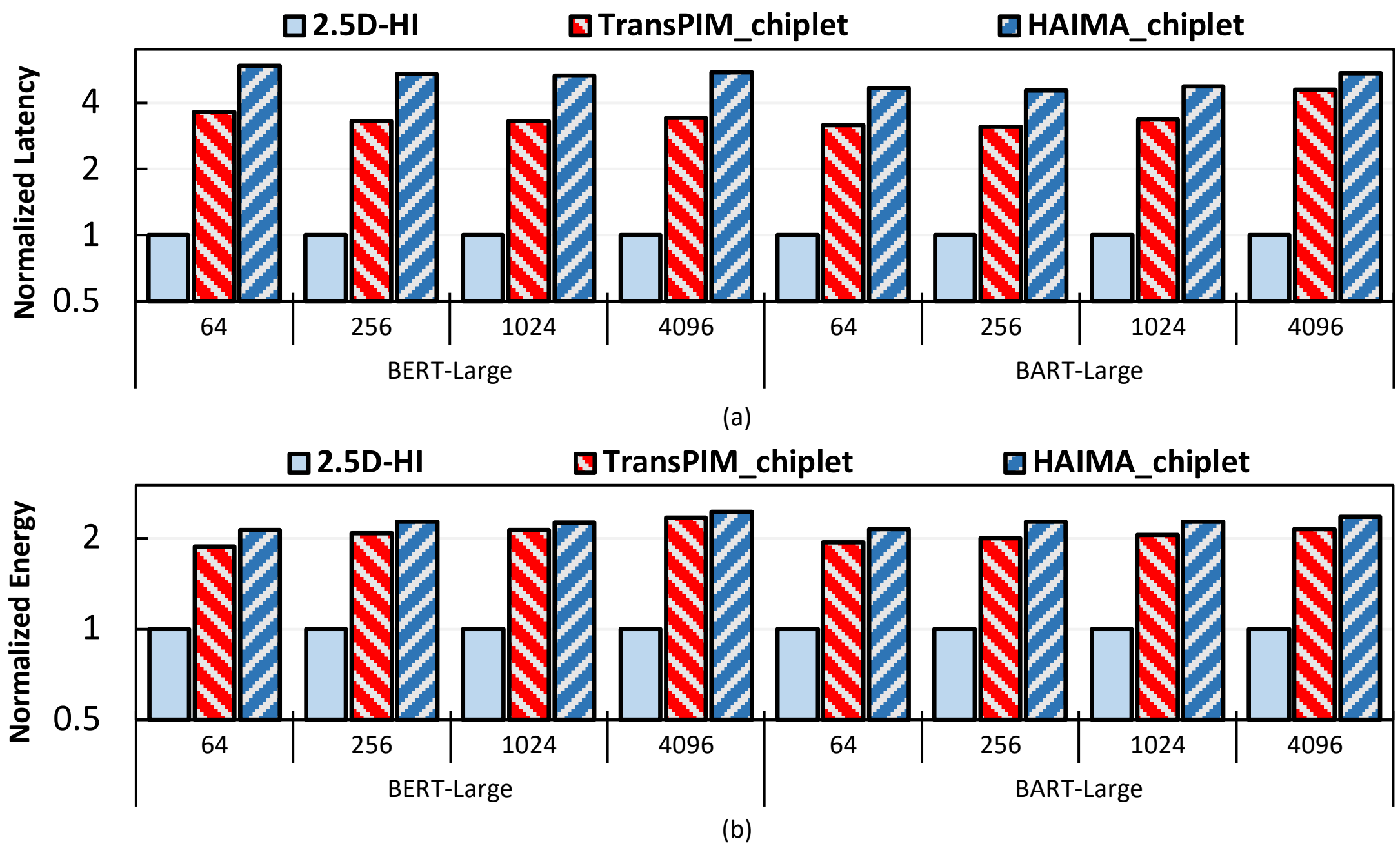

Figs. 9: End-to-end latency and energy improvement of the 2.5D-HI over its counterparts for 64 chiplet system running BERT-Large and BART-Large.

calculations on the SM chiplets. However, HAIMA and TransPIM use additional host access, which prevents online execution as it results in repeated data exchange with the host and hence incurs additional communication latency. Although HAIMA outperforms TransPIM in score computation, TransPIM has faster execution and lower energy consumption since it performs the FF network more efficiently. It should be noted that the implementation of SFC-based NoI for the feed-forward layers in TransPIM_chiplet and HAIMA_chiplet is not feasible due to non-deterministic traffic and continuous memory access dependencies, preventing the isolation of a dedicated sub-system exclusively for the FF layer. With HI, this constraint is mitigated by having a dedicated computational platform (ReRAM chiplets) to enable FF network processing.

We do not consider the ReRAM-only architecture, ReTransformer in this comparative performance evaluation for two reasons [1]. First, the ReTransformer architecture only focuses on speeding up self-attention with no full system evaluation. Second, the ReRAM-only architecture has endurance and reliability issues when accelerating Transformer models. For example, consider the BERT model with $h = 8$ heads and sequence length $n = 4096$. In this case, for the ReRAM chiplet, each with 16 tiles with 40 crossbar arrays of 128X128 storing 2 bits, we have 5KB of storage for a single write. In the case of self-attention specifically, $W_K, W_Q, W_V$ (512X64 matrices with each element represented in a 16-bit precision) and the $K, Q, V$ computation generates about $10^7$ writes (~30Gb) per ReRAM cell per token. For longer sequence lengths such as $N$=4096, the rewrites increase to $10^{10}$ in a single encoder. In the case of subsequent $Score, P_i$, and $H^{MHA}$, writes are of the order $10^7$. As the models grow following previous trends (GPT4, Llama2-70B with 170B parameters), the write updates would invoke serious endurance and reliability challenges for a ReRAM-only architecture [30]. 2.5D-HI avoids this challenge by integrating heterogenous chiplets on the 2.5D interposer and accelerating only the static parts (i.e., no write updates) of the computation using ReRAM chiplets. Hence, a heterogenous integration approach is imperative for accelerating Transformer models.

### 4.2 System Scalability of the Heterogenous Architecture

We extend our framework to larger models on 64 and 100 chiplet systems to demonstrate system scalability. Figs. 9(a) & (b) presents a comparative analysis showing full system (end-to-end) latency and energy for 2.5D-HI, TransPIM_chiplet, and HAIMA_chiplet running BERT-Large and BART-Large on a 64-chiplet system. We observe that 2.5D-HI consistently outperforms both TransPIM_chiplet and HAIMA_chiplet in terms of latency as well as energy consumption for all considered sequence lengths. Notably, the performance gain for 2.5D-HI increases with the input sequence length, showing scalability over sequence length. For instance, the latency gains increases from 4.6x to 5.45x as the input sequence length increases from 64 to 4096 for BART-Large. Similarly, energy gain for 2.5D-HI also increases as the sequence length increases with respect to HAIMA_chiplet and TransPIM chiplet. This performance enhancement is primarily due to heterogenous integration enabled through 2.5D-HI, which ensures each computational kernel of the Transformer model is executed on the appropriate hardware platform. Further, long-range communication is minimized on the 2.5D-HI system. Unlike HAIMA_chiplet and TransPIM_chiplet, the Softmax computation is performed online on the SM cores without necessitating any computation on the host device leading to communication bottlenecks. Further, during the execution of the feed-forward layer, the entire data flow is confined within the ReRAM macro, interconnected as a SFC. This minimizes contention issues due to limited data access outside of the ReRAM macro to other chiplets.

Next, we extend our system to 100 chiplet configurations and evaluate bigger transformer models featuring billions of parameters. Figs. 10(a) & (b) show the end-to-end latency and energy for Llama2 and GPT-J on the 100 chiplet system. We also consider the original TransPIM and HAIMA architectures to show the overall trend. We note up to 11.8× gain in latency and 2.36× lower energy consumption with respect to our baselines HAIMA_chiplet and TransPIM_chiplet. Moreover, we observe that the performance gain is up to 38x higher than the original TransPIM and HAIMA implementation. This is attributed to the limited number of banks that can be activated in parallel in the original 3D architecture due to the thermal challenges discussed in the subsection below.

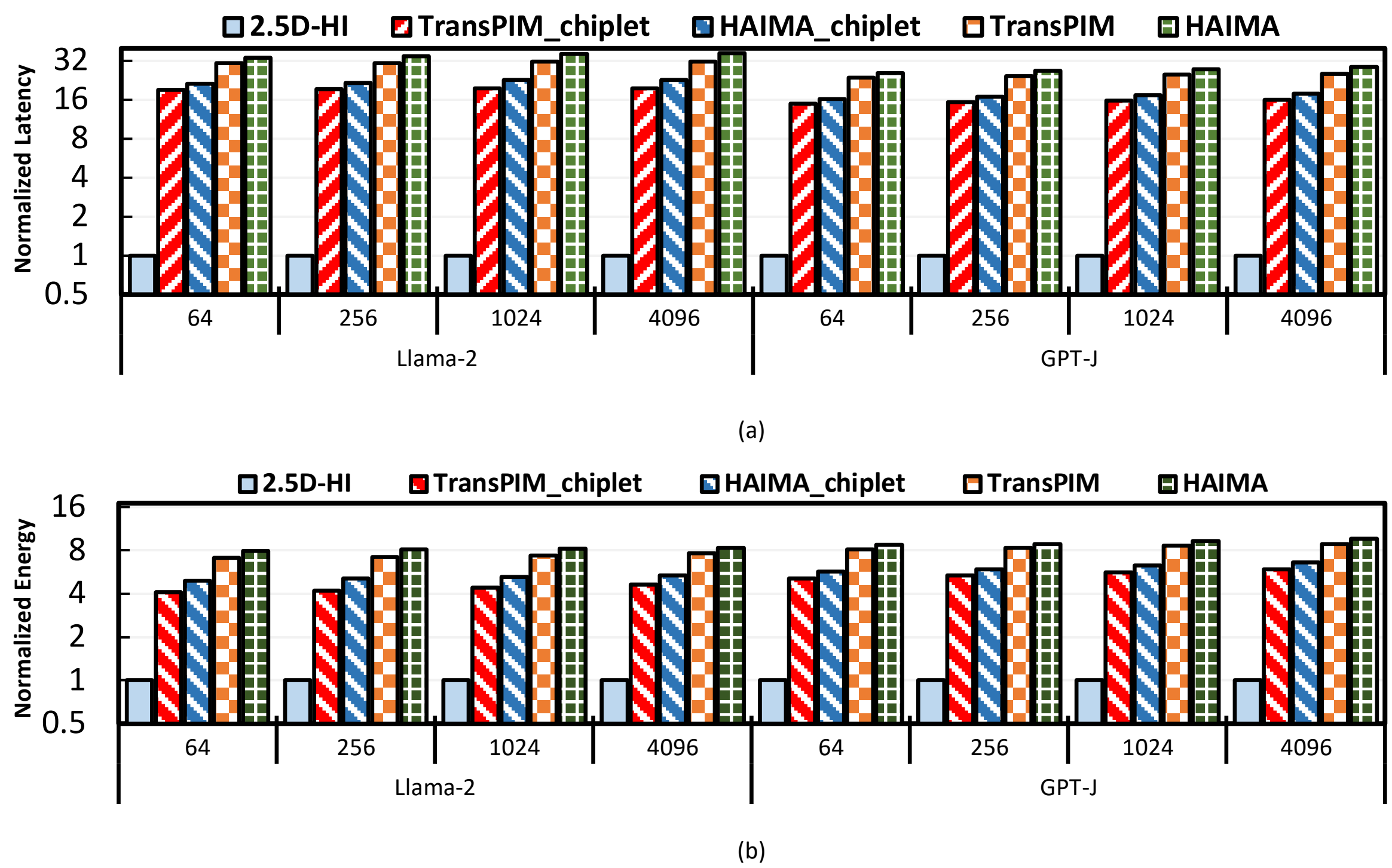

Fig. 10: End-to-end latency and energy improvements of the 2.5D-HI over its counterparts for 100 chiplet system running Llama2-7B and GPT-J.

Moreover, Table 4 (a) & (b) shows the end-to-end execution time for the 36-chiplet system executing BERT-Base (for n = 64 tokens) & 100-chiplet system executing GPT-J (for n = 64 tokens) respectively. It can be observed that 2.5D-HI consistently outperforms all its counterparts. 2.5D-HI architecture demonstrates *both model scalability and sequence scalability*, exhibiting faster execution as the transformer model size and input sequence length increase. In contrast, HAIMA_chiplet and TransPIM_chiplet face scalability challenges when accommodating larger model sizes. This is attributed to the increased communication overhead for both TransPIM_chiplet and HAIMA_chiplet, with the number of hops increasing as the system size grows due to long-range communication. These observations underscore the effectiveness of heterogenous integration and the NoI design in addressing the demands of larger-scale Transformer models, exhibiting superior scalability and performance.

### 4.3 Three-dimensional Heterogeneous Integration

We now explore three-dimensional (3D) heterogeneous integration (3D-HI) for transformer model acceleration, where planar tiers are vertically integrated with each tier consisting of multiple ReRAM or SM-MC chiplets. This 3D integration approach brings chiplets closer and helps reduce inter-chiplet data transfer costs. Further, vertical links provide an additional degree of freedom for data exchange, which helps to reduce congestion in many-to-few traffic scenarios detailed in Section 3. Fig. 2 illustrates a conceptual SM-MC and ReRAM 3D-HI architecture consisting of vertically integrated tiers, each with either ReRAM or SRAM chiplets. Note that both the SM-MC chiplets and ReRAM chiplets cannot be integrated on the same tier due to technology limitations. The planar tiers are connected vertically with TSV-based links [56] [57] [58].

Table 4: Absolute time of execution for different architectural configuration for (a) 36-chiplet system executing BERT-Base for n = 64 tokens; (b) 100-chiplet system executing GPT-J for n = 64 tokens. Reported time is in milliseconds (ms).

| Time Execution (ms) | 36 chiplets | 100 chiplets |
|---|---|---|
| TransPIM_chiplet | 210ms | 1435ms |
| HAIMA_chiplet | 340ms | 975ms |
| 2.5D-HI | 50ms | 143ms |

While 3D integration offers significant advantages, it also increases power density, leading to thermal hotspots. Consequently, it becomes necessary to consider the peak temperature of the system when optimizing for performance in 3D systems. Our existing MOO framework presented in Eq. 10 does not consider thermal as an objective and focuses on maximizing performance. Hence, we need additional design objectives beyond those listed in Eq. 11 through Eq. 13. Hence, we include additional design objectives for 3D-HI.

The peak temperature of each core can be estimated using the approximate thermal model, which determines the temperature using horizontal and vertical heat flow [59]. The vertical heat flow can be determined by dividing the system into vertical columns. The temperature of a core located at layer $k$ from the sink due to the vertical heat flow is given by:

$$T(n,k) = \sum_{i=1}^{k}\left(P_{n,i}\sum_{j=1}^{i} R_j\right) + R_b \sum_{i=1}^{k} P_{n,i} \tag{16}$$

where $P_{n,i}(t)$ is the power consumption of the core at layer $i$ from the sink in the vertical column $n$, $R_j$ is the thermal resistance in the vertical direction, and $R_b$ is the thermal resistance of the base layer [59].
The horizontal heat flow is represented through the maximum temperature difference in a layer.

$$\Delta T(k) = \max_{\mathrm{n}} T_{n,k} - \min_{\mathrm{n}} T_{n,k} \tag{17}$$

Combining both horizontal and vertical heat flow models, the objective becomes minimizing the worst-case temperature.

$$T(\lambda) = \left(\max_{\mathrm{n,k}} \{T(n,k)\}\right)\left(\max_{\mathrm{k}} \{\Delta T(k)\}\right) \tag{18}$$

Furthermore, 3D integration with heterogeneous chiplets introduces additional design complexities in comparison to the homogeneous counterpart. Specifically, ReRAM chiplets have higher thermal susceptibility when compared to the CMOS-based SM-MC chiplets [60]. ReRAM cells store weights as conductance, which is temperature-dependent, making them susceptible to thermal fluctuations and noise [61]. This can lead to erroneous computations and potentially degrade the inference accuracy of the underlying model [62]. The thermal objective defined in Eq. 18 does not capture the effect of non-idealities on ReRAM cells [63] [64]. Hence, we incorporate an additional objective to capture sensitivity of ReRAM core to thermal noise. The effect of thermal noise on the ReRAM conductance value can be represented by the following.

$$Noise(\lambda) = N\left(0, \frac{\sqrt{4G.K_b.T_{ReRAM}.F}}{V}\right) \tag{19}$$

Here $G$ represents ReRAM ideal conductance, $K_b$ is the Boltzmann constant, $F$ denotes the operating frequency, and $V$ is the voltage difference across the two ReRAM terminals [62] [65]. We consider the temperature of ReRAM chiplets in the context of the ReRAM noise as the optimization objective.

Considering the performance, thermal, and ReRAM noise objectives, the updated MOO formulations for the design of 3D-HI is as follows:

$$\lambda^* = MOO\big(objectives = \mu(\lambda), \sigma(\lambda), T(\lambda), Noise(\lambda)\big) \tag{20}$$

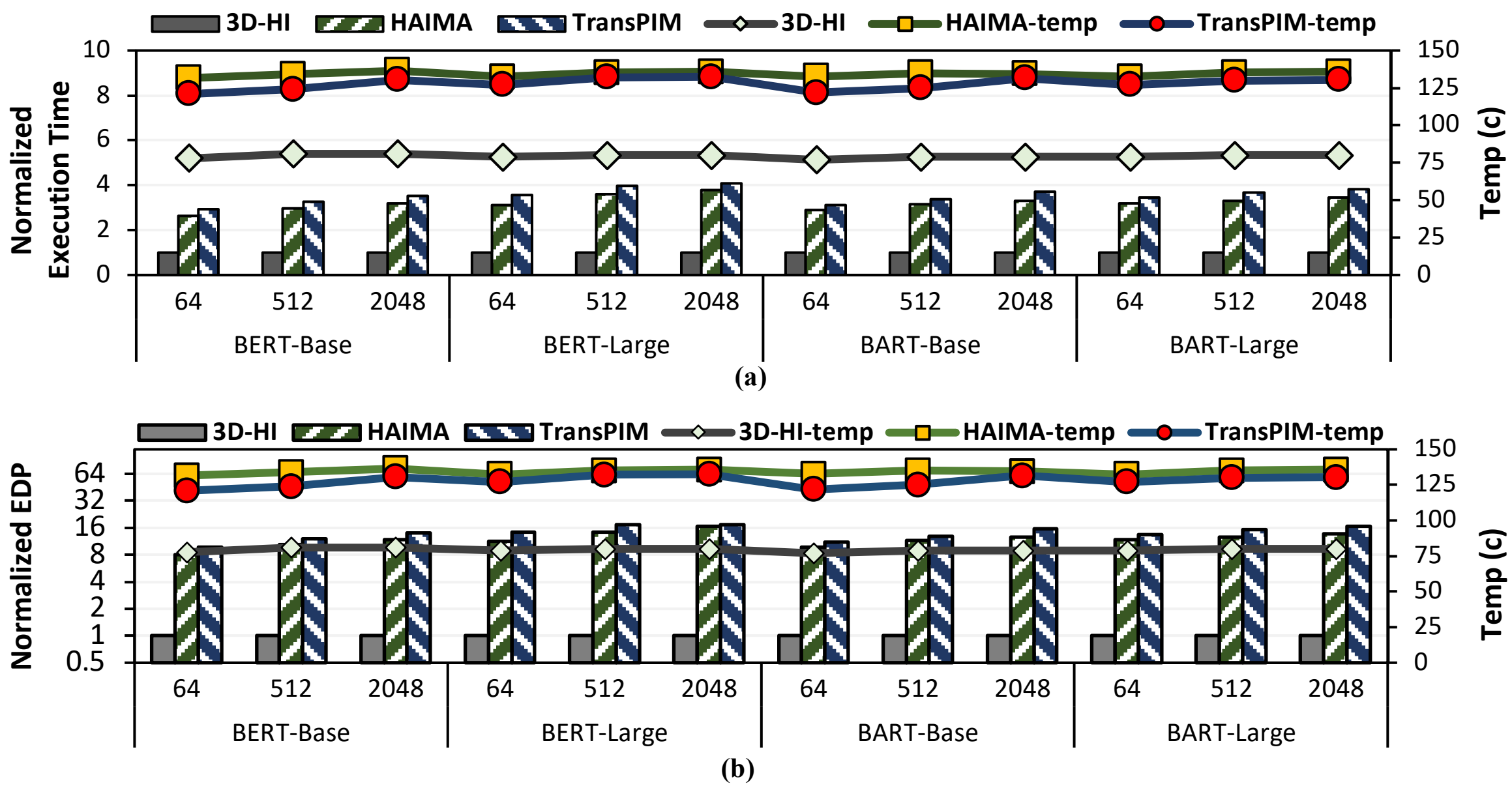

Fig. 11: Normalized (a) execution time and temperature, and (b) EDP and temperature, when compared to the baseline 3D-HI for different transformer models with varying sequence length n.

We now comapre end-to-end latency and energy of 3D-HI with the baslines HAIMA and TransPIM. However, unlike 3D-HI, where we perform joint performance-thermal-noise optimization, HAIMA and TransPIM did not consider thermal effects while designing the architecture. For instance, HAIMA's configuration suggests integration of up to eight compute units per bank, with each compute unit dissipating 3.138W of power. Under the assumption that multiple compute units in a bank are operating concurrently as suggested in HAIMA, the power density of the HBM bank will be order of magniture higher than modern GPUs given the standard HBM2 die area of 53.15 $mm^2$. Similarly, TransPIM consists of 8 stacks of HBMs connected through TSV. The thermal resistance increases as we move up in the stack and away from the heat sink. Consequently, additional power dissipation through compute units in banks can increase DRAM temperature quickly and create hotspots. Hence, we show system temperature along with execution and energy in our evaluation.

Fig. 11(a) presents the overall performance as well as steady state system temperature for real-world transformer models and input sequence lengths to conduct a thorough performance and thermal analysis. Notably, we observe that both HAIMA and TransPIM exhibit elevated temperatures, with a minimum temp of 120°C for any transformer model. The maximum temperature reaches 131°C in the case of the fused MHA-FF model where MHA and FF layers are computed parallelly. Note that the maximum temperature threshold for DRAM is 95°C beyond which it results in data loss or corruption. These results underscore the thermal infeasibility of both HAIMA and TransPIM, necessitating the exploration of dynamic voltage frequency scaling. However, exploring such techniques fall beyond the scope of the current work. Conversely, 3D-HI proves to be thermally realizable, consistently achieving speedup across various transformer models and sizes. Fig. 11(b) illustrates the normalized EDP along with steady state temperature for the transformer models and input sequence lengths considered above. We consider EDP as the relevant metric as it captures both performance and energy in a single term. 3D-HI outperforms the baselines, exhibiting increased EDP gains as the transformer model size and input sequence length increase, thereby showing scalability. For instance, the EDP of 3D-HI is an order of magnitude better (14.5×) than HAIMA for the BERT-Large model with n = 2056. In summary, 3D-HI demonstrates superior performance and temperature characteristics compared to the baselines.

## 5 CONCLUSION

The end-to-end transformer model exhibits significant heterogeneity in its computational kernels, necessitating the integration of different types of hardware modules on a single system for high-performance and energy-efficient acceleration. This paper considers different heterogeneous chiplets to design a multi-chiplet architecture called 2.5D-HI for accelerating Transformer models. 2.5D-HI uses SM-MC chiplets for multi-head attention and ReRAM chiplets for the feed-forward network, which optimize both achievable energy efficiency and throughput. The heart of the 2.5D-HI platform is a NoI architecture that enables efficient computing kernel to chiplet mapping for the complex interactions among heterogenous chiplets. Further, vertical integration on top of a 2.5D interposer helps to enhance overall system performance and alleviates the issue of memory bottlenecks. Experimental results demonstrate that 2.5D-HI lowers the latency and energy consumption by up to 11.8× and 2.36× with respect to an equivalent state-of-the-art chiplet-based platform. Notably, 2.5D-HI exhibits versatility across various transformer models/sizes and shows consistent speedup irrespective of the model.